\documentclass[preprint,prd,noshowpacs,nofootinbib]{revtex4}

\usepackage{color} 
\usepackage{amssymb}
\usepackage{amsmath}
\usepackage{graphicx} 
\usepackage{float}
\usepackage{braket}     
\usepackage{pdfpages}
\usepackage{epstopdf}
\usepackage[utf8]{inputenc}
\usepackage{soul}
\usepackage{enumitem}

\begin{document}

\title{Minimal Lengths in 3D via the Generalized Uncertainty Principle}

\author{Michael Bishop}
\email{mibishop@mail.fresnostate.edu}
\affiliation{Mathematics Department, California State University Fresno, Fresno, CA 93740}

\author{Joey Contreras}
\email{mkfetch@mail.fresnostate.edu}
\affiliation{Physics Department, California State University Fresno, Fresno, CA 93740}

\author{Peter Martin}
\email{kotor2@mail.fresnostate.edu}
\affiliation{Physics Department, California State University Fresno, Fresno, CA 93740}

\author{Piero Nicolini}
\email{piero.nicolini@units.it}
\affiliation{Dipartimento di Fisica, Università degli Studi di Trieste, 
Trieste, Italy \\ Frankfurt Institute for Advanced Studies (FIAS),
Frankfurt am Main, Germany \\
Institut f{\"u}r Theoretische Physik,
Johann Wolfgang Goethe-Universit{\"a}t-Frankfurt am Main,
Frankfurt am Main, Germany
}

\author{Douglas Singleton}
\email{dougs@mail.fresnostate.edu}
\affiliation{Physics Department, California State University Fresno, Fresno, CA 93740}
\affiliation{Kavli Institute for Theoretical Physics, University of California Santa Barbara, Santa Barbara, CA 93106, USA}

\date{\today}

\begin{abstract}
We investigate an extension of the Generalized Uncertainty Principle (GUP) in three dimensions by modifying the three dimensional position and momentum operators in a manner that remains coordinate-independent and retains as much of the standard position-momentum commutators as possible.  
Moreover, we bound the physical momentum which leads to an effective minimal length in every coordinate direction.  
The physical consequences of these modified operators are explored in two scenarios: (i) when a spherically-symmetric wave function is `compressed' into the smallest possible volume; (ii) when the momentum is directed in a single direction.  
In case (ii), we find that the three dimensional GUP exhibits interesting phenomena that do not occur in one dimension: the minimal distance in the direction parallel to a particle's momentum is different from the minimal distance in the orthogonal directions. 
 
\end{abstract}

\maketitle

\section{Introduction}
The Generalized Uncertainty Principle (GUP) and its associated minimal length scale, or maximum energy-momentum scale, are considered fruitful phenomenological approaches to quantum gravity \cite{vene, gross,amati2,amati,maggiore,garay,KMM,scardigli,adler-1999}. The GUP approach to quantum gravity has led to testable predictions and observations, {\it e.g.} explanations for photon dispersion in vacuum \cite{AC-nature} 
and in various table-top laboratory experiments \cite{vagenas,vagenas2}. While many works mention the three dimensional generalization of the position-momentum commutators and the associated three dimensional uncertainty relationships, they generally restrict their GUPs to one spatial dimension. There are a few exceptions including \cite{KMM} and more recently the work in \cite{todrinov}, which looked at a fully relativistic GUP in four space-time dimensions. 

In this work, we want to explore the GUP in three spatial dimensions and note some new features which are present in three spatial dimensions, but are absent in a single spatial dimension. 
This is similar to the theory of special relativity where boosts were introduced in only one spatial dimension and later expanded to three dimensions.  
Looking at special relativity in three spatial dimensions leads not only to a more complex form for the boosts, but also gives rise to the mixing of boost transformations, rotations, and related effects such as Thomson precession. 
In a similar way to Lorentz boosts, we find that certain variations of three dimensional GUPs lead to an interplay between the different coordinate directions, resulting in different minimal distances in the directions orthogonal to the direction of momentum. 

It is interesting to notice that a similar behavior has already been observed in the context of string inspired noncommutative geometry \cite{carroll,smailagic,spallucci}. Noncommutative relations determine a foliation of the spacetime in noncommutative planes. If, on the one hand, the foliation is an invariant character, on the other hand, on each of such planes the minimal length transforms like a antisymmetric second rank tensor, {\it i.e.}, the Kalb-Ramond field -- see {\it e.g.} \cite{seiberg}. 

\section{Minimal Lengths in one dimension}

In this section we will review the general construction of a one dimensional GUP, and introduce a specific form of the GUP that we will focus on in this work. The specific GUP studied here  will modify the position and momentum operators, while {\it trying} to retain the standard commutator between conventional position and momentum.

Most works on GUP start with a modified commutator  of the form
\begin{equation}
[{\hat X} , {\hat P}] = i \hbar F\left(\frac{{\hat x}}{x_m}, \frac{{\hat p}}{p_M} \right),
\label{eq:modcomm}
\end{equation}
where $F\left(\frac{{\hat x}}{x_m}, \frac{{\hat p}}{p_M} \right)$ is a function of the canonical position and momentum operators which satisfy $[{\hat x} , {\hat p}] = i \hbar $; and the capitalization on $\hat{X}$ and $\hat{P}$ is used to denote the modified operators. 
The constants $x_m$ ($p_M$) set the small distance (large momentum) cut-off scale. 
When $\Delta x \gg x_m$ and/or $\Delta p \ll p_M$, the ansatz function has the limit $ \langle F\left(\frac{{\hat x}}{x_m}, \frac{{\hat p}}{p_M} \right) \rangle \to 1$, and the modified commutator becomes the standard one for large distances and/or small momenta. A common example of such a modified commutator is one from \cite{KMM} where $F\left(\frac{{\hat x}}{x_m}, \frac{{\hat p}}{p_M} \right) = 1 + \beta {\hat p}^2 = 1 + \frac{{\hat p}^2}{p_M ^2}$, with $\beta \equiv p_M ^{-2}$, to connect our notation with that of \cite{KMM}.

With a modified commutator defined, it is now necessary to find corresponding modified quantum operators that lead to that modified commutator.  
An acceptable modification would be to modify the operators in momentum space to have the form
\begin{equation}
    \label{xp1d}
{\hat X} = i \hbar G\left(\frac{{\hat x}}{x_m}, \frac{{\hat p}}{p_M} \right) \partial _p  ~~~~ {\rm and} ~~~~ {\hat P} =  H\left(\frac{{\hat x}}{x_m}, \frac{{\hat p}}{p_M} \right) ~{\hat p}{}~,
\end{equation}
and choose $G$ and $H$ so as to yield  \eqref{eq:modcomm}.

For example, with $F\left(\frac{{\hat x}}{x_m}, \frac{{\hat p}}{p_M} \right) = 1 + \beta {\hat p}^2$ the operators can be defined as  ${\hat X} = i \hbar (1 + \beta p^2) \partial _p$ and ${\hat P} = {\hat p} = p$. Here, the function $F$ depends only on the momentum and not the position, and the momentum operator is simply the canonical momentum operator {\it i.e.} multiplication by $p$. From here, we will take this specialized case where the ansatz functions modify the position and momentum operators which leads to a commutator that depends only on the canonical momentum, $p$.\footnote{Also from this point onward we will not be overly careful about distinguishing ${\hat p}$ from the momentum input variable $p$ since we consider these operators as acting on wave functions defined in momentum space, i.e. $\psi(p)$.}
Another acceptable set of modified operators which gives the same modified commutator from \eqref{eq:modcomm} is  ${\hat X} = {\hat x} = i \hbar \partial _p$ and ${\hat P} = p  \left(1 + \frac{1}{3} \beta p^2 \right)$,  

In \cite{aiken-2019,BLS} it was shown that there are a host of different ways to modify the position and momentum operators that lead to the same modified commutator. However, in some cases these different modifications of the operators, while giving the same modified commutator, did not lead to a minimum distance. For this reason, the works \cite{BLS,BJLS} assert that it is primarily the form of modification of the operators, rather than the form of the modified commutator, that determines a minimal length scale in GUP models. This contrasts with the more common approach of first choosing the modified commutator and then choosing the (non-unique) modified operators which give this modified commutator. 

Reference \cite{BJLS} takes the ansatz functions $G$ and $H$ in \eqref{xp1d} such that ${\hat X}$ and ${\hat P}$ satisfied the standard commutator relationship
\begin{equation}
[{\hat X}, {\hat P}]=i \hbar.   
\label{eq:cancomm-modvar}
\end{equation} 
$H$ was chosen such that when $p  \ll p_M$, $\hat{P}=p ~ H\left( \frac{p}{p_M} \right) \approx p$;
and such that when $p \gg p_M$, $\hat{P}= p ~H\left( \frac{p}{p_M} \right) \approx p_M$, t cap momentum at $p_M$. 

For the ansatz function $G\left( \frac{p}{p_M} \right)$, we require $G\left( \frac{ p}{p_M} \right) \to 1$ for $\Delta  p \ll p_M$. The limit of $G\left( \frac{p}{p_M} \right)$ for $p \gg p_M$ is fixed below by requiring that the modified position and momentum operators still satisfy the standard commutator \eqref{eq:cancomm-modvar}. 
These choices for the ansatz functions still give a minimum in position because of the following: 
\begin{enumerate}[label=(\roman*)]
    \item The general form of the ansatz function $H$ gives ${\hat P} \le p_M$.
    \item Since  $G\left( \frac{p}{p_M} \right)$ and $H\left( \frac{p}{p_M} \right)$ are fixed to maintain $[{\hat X}, {\hat P}]=i \hbar$, this yields the standard uncertainty principle $\Delta X \Delta P \ge \frac{\hbar}{2}$.
    \item Finally,  this leads to a minimum uncertainty in position, $\Delta X  \ge \frac{\hbar}{2 p_M}$. 
\end{enumerate}
Now, one obtains a condition on $G$ and $H$, 
\begin{equation}
    \label{cond1}
    G\left( \frac{p}{p_M} \right) \partial_p \left[ H\left(\frac{p}{p_M} \right)~p\right] =1. 
\end{equation}
In \cite{BJLS}, two specific sets of ansatz functions where given which satisfied the conditions above,
\begin{equation}
    \label{xp1da}
    G\left( \frac{p}{p_M} \right) = \cosh ^2 \left( \frac{p}{p_M}\right) ~~~~{\rm and}~~~~
    H\left( \frac{p}{p_M} \right) = \frac{p_M}{p} \tanh \left( \frac{p}{p_M} \right) ~,
\end{equation}
and 
\begin{equation}
    \label{xp1db}
    G\left( \frac{p}{p_M} \right) = \left[ 1+ \left( \frac{\pi p}{2 p_M}\right)^2 \right] 
    ~~~~{\rm and}~~~~
    H\left( \frac{p}{p_M} \right) = \frac{2 p_M}{\pi p} \arctan \left( \frac{\pi p}{2 p_M} \right) ~.
\end{equation}
The ansatz functions in \eqref{xp1da} and \eqref{xp1db} satisfy condition \eqref{cond1}. 

In both cases, the modified operators, ${\hat X}$ and ${\hat P}$, formulated by using the $G$'s and $H$'s in \eqref{xp1d} satisfy the standard commutator $[{\hat X}, {\hat P}] = i \hbar$, and the related standard uncertainty $\Delta X \Delta P \ge \frac{\hbar}{2}$. One might think recovering the standard commutator  \eqref{eq:cancomm-modvar} would not lead to a minimum length, but due to the modified behavior of ${\hat P}$, and therefore $\Delta P$, one still finds a non-zero minimum distance. 
In conclusion, the crux of the question is if operators are capped or not. Physical operators must display ultraviolet convergence/finiteness. Uncapped operators have only a mathematical meaning, although can be used as auxiliary variables in calculations. The algebra of operators is per se irrelevant to discriminate the short distance nature of observables. 

\section{Generalizing to Three Dimensions}

In this section, we generalize the above construction from one spatial dimension to three spatial dimensions. The modified operators must now become three-vector operators -- ${\hat X} \to {\hat X}_i$ and ${\hat P} \to {\hat P} _i$, where $i=1,2,3$. This can be done by applying different conditions to the ansatz functions, $G$ and $H$ as we now discuss in detail. 

For the hyperbolic ansatz functions of \eqref{xp1da}, one could try letting $H  \left( \frac{p}{p_M} \right) = \frac{p_M}{p_i} \tanh \left( \frac{p_i}{p_M} \right)$, which gives a modified momentum of the form ${\hat P}_i = p_i H  \left( \frac{p}{p_M} \right) = p_M \tanh \left( \frac{p_i}{p_M} \right)$. This would have a low momentum limit of ${\hat P}_i \approx p_i$, and a high momentum limit of $p_M$ in the direction of $p_i$. The vector character of the modified position operator arises through $\partial_p \to \partial_{p_i}$, and the ansatz function becomes $G \left( \frac{p_i}{p_M} \right) = \cosh ^2 \left( \frac{p_i}{p_M}\right)$.
In order to satisfy \eqref{cond1}, the $\cosh ^2$ term for the modified position operator has to have the same functional dependence as the $\tanh$ term in the modified momentum operator. Note that condition \eqref{cond1} becomes $G\left( \frac{p_i}{p_M} \right) \partial_{p_i} \left[ p_j H\left(\frac{p_j}{p_M} \right)\right] = \delta _{ij} $ in this three dimensional case.  

This approach has a fundamental problem, though. $G$ and $H$ are functions of $p_i$, which spoils the rotational symmetry of the modified position and momentum operators, as discussed in section 6.2 of \cite{KMM}. 

In order to preserve rotational symmetry and still retain the feature that the momentum operators have upper bounds of the form given in section II, we need ansatz functions of the form
\begin{equation}
    \label{xp3da}
    G  \left( \frac{|{\vec p}|}{p_M} \right) = \cosh ^2 \left( \frac{|{\vec p}|}{p_M}\right) ~~~~{\rm and}~~~~
    H \left( \frac{|{\vec p}|}{p_M} \right) =  \frac{p_M}{|{\vec p}|} \tanh \left( \frac{|{\vec p}|}{p_M} \right) ~,
\end{equation}
or for the arctan modification
\begin{equation}
    \label{xp3db}
    G  \left( \frac{|{\vec p}|}{p_M} \right) = \left[ 1+ \left( \frac{\pi |{\vec p}|}{2 p_M}\right)^2 \right] 
    ~~~~{\rm and}~~~~
    H \left( \frac{|{\vec p}|}{p_M} \right) = \frac{2 p_M}{\pi |{\vec p}|} \arctan \left( \frac{\pi |{\vec p}|}{2 p_M} \right) ~.
\end{equation}
These ansatz functions in \eqref{xp3da} and \eqref{xp3db} lead to modified three dimensional position and momentum operators of the form
\begin{equation}
    \label{xp3dc}
    {\hat X}_i = i \hbar \cosh ^2 \left( \frac{|{\vec p}|}{p_M}\right) \partial_{p_i} ~~~~{\rm and}~~~~
     {\hat P}_i=  p_i~H \left( \frac{|{\vec p}|}{p_M} \right) = p_i \frac{p_M}{|{\vec p}|} \tanh \left( \frac{|{\vec p}|}{p_M} \right) ~,
\end{equation}
or for the arctan modification
\begin{equation}
    \label{xp3dd}
    {\hat X}_i = i \hbar \left[ 1+ \left( \frac{\pi |{\vec p}|}{2 p_M}\right)^2 \right] \partial_{p_i} 
    ~~~~{\rm and}~~~~
    {\hat P}_i = p_i~H \left( \frac{|{\vec p}|}{p_M} \right) = p_i \frac{2 p_M}{\pi |{\vec p}|} \arctan \left( \frac{\pi |{\vec p}|}{2 p_M} \right) ~.
\end{equation} The modified position  and momentum operators in \eqref{xp3dc} and \eqref{xp3dd} are the three dimensional generalizations of \eqref{xp1da} and \eqref{xp1db}, which satisfy rotational symmetry and still have capped momentum.

We will now show that the position and momentum operators' dependence on $|{\vec p}|$ causes the standard commutator ({\it i.e.}, $[{\hat X}_i , {\hat P}_j] = i \hbar \delta _{ij}$) to no longer necessarily be maintained in all directions. The standard commutator can be maintained in one direction but at the cost of altering the commutator in the two orthogonal directions, resulting in there being different minimal distances in different directions. In one dimension, the condition for the commutator to retain its standard form was given by \eqref{cond1}. To generalize \eqref{cond1} to three dimensions one would like to require
\begin{equation}
\label{cond3d}
    G  \left( \frac{|{\vec p}|}{p_M} \right) \partial_{p_i} \left[ p_j ~H \left(\frac{|{\vec p}|}{p_M} \right)\right]  \stackrel{?}{=} \delta _{ij} ~,
\end{equation}
where the question mark indicates we are asking if it is possible for the left hand side to equal the right hand side. 

The complication that arises with \eqref{cond3d}, that is not found in \eqref{cond1}, is that the momentum derivative operates on all the components of momentum ``hidden" in the function in square brackets due to the $|{\vec p}|$ dependence of $H$. This leads to $p_i$ and $p_j$ terms that are not found in the one dimensional case. 

For instance, for the $\tanh$ modification of the momentum, condition \eqref{cond3d} becomes
\begin{eqnarray}
    \label{tanh-cond}
    G  \left( \frac{|{\vec p}|}{p_M} \right) \partial_{p_i} \left[ p_j ~H \left(\frac{|{\vec p}|}{p_M} \right)\right] &=& \left[ \frac{p_M}{|{\vec p}|} \sinh \left( \frac{|{\vec p}|}{p_M} \right) \cosh \left( \frac{|{\vec p}|}{p_M} \right)\left( \delta_{ij} - \frac{p_j p_i}{|{\vec p}|^2} \right) + \frac{p_j p_i}{|{\vec p}|^2} \right] \nonumber \\ &=& \delta_{ij} + \mathcal{O}\left(\frac{|\vec{p}|^2}{p_M^2} \right) ~,
\end{eqnarray}
and the $\arctan$ modification of the momentum, condition \eqref{cond3d} becomes
\begin{eqnarray}
    \label{arctan-cond}
    G  \left( \frac{|{\vec p}|}{p_M} \right) \partial_{p_i} \left[ p_j ~H \left(\frac{|{\vec p}|}{p_M} \right)\right] &=& \left[ \left( 1+ \left( \frac{\pi |{\vec p}|}{2 p_M}\right)^2 \right) \frac{2 p_M}{\pi |{\vec p}|} \arctan \left( \frac{\pi |{\vec p}|}{2 p_M} \right)\left( \delta_{ij} - \frac{p_j p_i}{|{\vec p}|^2} \right) + \frac{p_j p_i}{|{\vec p}|^2} \right] \nonumber \\
    &=& \delta_{ij} + \mathcal{O}\left(\frac{|\vec{p}|^2}{p_M^2} \right) ~.
\end{eqnarray}
In the second lines of \eqref{tanh-cond} and \eqref{arctan-cond}, we expanded the functions in square brackets on the right hand side to first order in $|{\vec p}|/p_M$. To first order, one can recover the standard commutators in three dimensions.  However, to second order, there is a difference with the one dimensional case, and the expressions in \eqref{tanh-cond} and \eqref{arctan-cond} are no longer equal to solely $\delta_{ij}$.

Using the modified position and momentum from \eqref{xp3dc} and \eqref{xp3dd}, and the results of \eqref{tanh-cond} and \eqref{arctan-cond}, we find that the modified commutator for the $\tanh$ modification, to second order, is   
\begin{equation}
    \label{tanh-comm}
    [{\hat X}_i , {\hat P}_j] \approx i \hbar     
        \left[  \delta_{ij} + \frac{|{\vec p}|^2}{2 p_M ^2} \left( \delta_{ij} - \frac{p_j p_i}{|{\vec p}|^2} \right) \right]~.
\end{equation}

For the $\arctan$ modification there is a bit if a subtlety. From \eqref{arctan-cond}, the term we want to expand is of the form $(1+x^2) \frac{\arctan (x)}{x}$, with $x \equiv \frac{\pi |{\vec p}|}{2 p_M}$. Due to the $x$ in the denominator of this expression, one needs to expand $\arctan$ to ${\cal O} (x^3)$ namely $\arctan (x) \approx x - \frac{x^3}{3} $. In this way \eqref{arctan-cond} to ${\cal O} (x^2)$ is      
\begin{equation}
    \label{arctan-comm}
     [{\hat X}_i , {\hat P}_j] \approx i \hbar     \left[ \delta_{ij}+ \frac{2}{3} \left( \frac{\pi |{\vec p}|}{2 p_M}\right)^2  \left( \delta_{ij} - \frac{p_j p_i}{|{\vec p}|^2} \right)  \right] ~.
\end{equation}

The second order approximations in \eqref{tanh-comm} and \eqref{arctan-comm}, are similar to the modified commutators in \cite{KMM} in that the modifications are quadratic in $|{\vec p}|$. However, unlike the one dimensional case, one can no longer recover the standard commutator \eqref{eq:cancomm-modvar} with the modified operators from \eqref{xp3dc} and \eqref{xp3dd}.

\section{Physical consequences of three dimensional GUP}

We now want to investigate the physical consequences of the modified commutators of the previous section on the resulting uncertainty principle. The left hand side of the uncertainty principle will be the product of the variation of the modified position  and momentum operators, $\Delta X_i \Delta P_j$, where for a general operator ${\hat A}$ the variation is $\Delta A = \sqrt{\langle {\hat A}^2 \rangle -\langle {\hat A} \rangle ^2 }$. The right hand side of the uncertainty relationship involves taking the expectation of the right hand side of \eqref{tanh-cond} or \eqref{arctan-cond}. However, the presence of the hyperbolic functions in \eqref{tanh-cond} or the inverse tangent function in \eqref{arctan-cond} makes it hard to take the expectation values and get an analytical form. In order to more easily calculate the expectation values, we will instead use the second order commutators given in \eqref{tanh-comm} and \eqref{arctan-comm}. Even after this, we will need to make another approximation to obtain the final form of the uncertainty principle.

We will look at two different cases for evaluating the physical consequences of the 3D GUP: (a) We assume a spherically symmetric wave function so that the average of the modified momentum and position operators is zero so that $\Delta X_i= \sqrt{\langle {\hat X_i}^2 \rangle }$, and $\Delta P_j= \sqrt{\langle {\hat P_j}^2 \rangle }$. (b) We assume that the system has a non-zero, average momentum in one direction ({\it e.g.} the $i=1$ direction) so that $\langle {\hat P}_i \rangle \ne 0$). 

For both cases, the minimum distance will be proportional to the inverse of the cut-off scale $\Delta X ^{min} \propto \frac{1}{p_M}$,
up to numerical factors of ${\cal O} (1)$. For the second case, with an average momentum in one direction, we find that the minimum position uncertainty in the direction of the momentum is smaller than in the two perpendicular directions. 

\subsection{Spherically Symmetric Case}

In this subsection we study the case where the wave function is spherically symmetric in momentum space. We use this to explore how focused/concentrated a stationary particle can get in position space in these theories.
We we utilize the spherical symmetry to make the substitution $\langle p_j p_i \rangle = \delta _{ij} \langle p_i ^2 \rangle$.

Starting with the hyperbolic case \eqref{tanh-comm}, the modified uncertainty relationship is
\begin{equation}
    \label{tanh-gup}
    \Delta X_i \Delta P_j \ge \frac{\hbar}{2} \left(\delta_{ij} + \frac{\delta _{ij}}{2 p_M ^2} \langle |{\vec p}|^2 - p_i ^2 \rangle  \right)  = \frac{\hbar}{2} \delta_{ij} \left( 1 + \frac{\langle p_{\perp} ^2 \rangle}{2 p_M ^2}   \right) ~,
\end{equation}
where $\langle p_{\perp} ^2 \rangle = \langle |{\vec p}|^2 - p_i ^2 \rangle$ is the expectation of the square of the perpendicular part of the momentum.  
We can use the symmetry to write $\langle p_{\perp}\rangle^2= (d-1) \langle p_i ^2\rangle$, where $d$ is the number of spatial dimensions. In this case, $d=3$ and $|p_{\perp}|^2= 2 |p_i| ^2$ so \eqref{tanh-gup} becomes
\begin{equation}
    \label{tanh-gup-2}
    \Delta X_i \Delta P_j \ge   \frac{\hbar}{2} \delta_{ij} \left( 1 + \frac{\langle |p_i| ^2 \rangle}{p_M ^2}   \right) ~.
\end{equation}
Next, we need to estimate $\Delta P_j$,
\begin{equation}
    \label{dpj}
    \Delta P_j = \sqrt{\langle ({\hat P}_j)^2 \rangle - \langle {\hat P}_j\rangle ^2} \le  \sqrt{\langle ({\hat P}_j)^2 \rangle} =
    \sqrt{\left\langle \frac{{\hat p}_j^2 p_M^2}{|\vec p|^2}  \tanh ^2 \left( \frac{|{\vec p}|}{p_M} \right)\right\rangle} 
\end{equation}
where we take $\langle {\hat P}_j\rangle =0$; and bounding $\tanh ^2(x) \leq x^2$,  equation \eqref{dpj} yields
\begin{equation}
    \label{dpj2}
    \Delta P_j \le 
    \sqrt{\left\langle \hat p_j ^2 \right\rangle} = \Delta p_j  ~,
\end{equation}
where the uncertainty of the modified momentum is bound by the uncertainty of the canonical momentum. 
Combining \eqref{tanh-gup-2} and \eqref{dpj2} gives a $\Delta X_i$ of
\begin{equation}
    \label{tanh-up-2}
   \Delta X_i \ge   \frac{\hbar}{2} \left( \frac{1}{\Delta p_i} + \frac{\Delta p_i}{p_M ^2}   \right)~.
\end{equation}
The minimum of \eqref{tanh-up-2} occurs at $\Delta p_i = p_M$ which then gives
\begin{equation}
    \label{tanh-xmin}
    \Delta X^{min} _i = \frac{\hbar}{p_M}. 
\end{equation} 
The structure for this 3D GUP given by \eqref{tanh-gup-2} \eqref{tanh-up-2} and \eqref{tanh-xmin} is very similar to, but not exactly identical to, the original KMM model \cite{KMM}. 

One could also do a rougher, but simpler, approximation and get close to the same result as \eqref{tanh-xmin}: instead of \eqref{tanh-gup-2}, simply take $\Delta X_i \Delta P_j \ge   \frac{\hbar}{2} \delta_{ij}$; and instead of \eqref{dpj2} one could take $\Delta P_j \le p_M$. 
This yields $\Delta X_i ^{min} = \frac{\hbar}{2 p_M}$ which is of the same order as \eqref{tanh-xmin}.
There is not a big difference between this rough estimate for $\Delta X_i ^{min}$ and the more refined one in \eqref{tanh-xmin}.

We now repeat this analysis for the $\arctan$ modification given by \eqref{arctan-cond}. At first glance, it seems this is straight forward, but there is a subtle issue as compared to the previous $\tanh$ case. For the $\tanh$ modification, in going from \eqref{tanh-cond} to the approximation in \eqref{tanh-comm}, we note that the right hand side of \eqref{tanh-comm} is a lower bound to \eqref{tanh-cond}, thus in \eqref{tanh-gup} we are justified in using $\ge$. However since the expansion for $\arctan$ has alternating signs, the approximation in \eqref{arctan-comm} is not always a lower bound on \eqref{arctan-cond}. This makes getting the uncertainty relationship coming from \eqref{arctan-cond} more complicated. 

One could make a similar approximation to the rough approximations mentioned after \eqref{tanh-xmin}. Note that the $(1+x^2) \frac{\arctan (x)}{x}$ term from \eqref{arctan-cond} (here $x= \frac{\pi |{\vec p}|}{2 p_M}$) has a lower bound of $1$, {\it i.e.} $(1+x^2) \frac{\arctan (x)}{x} \ge 1$. Replacing this term in \eqref{arctan-cond} by $1$ gives $[X_i, P_j] = i \hbar \delta_{ij}$, which then gives $\Delta X_i \Delta P_j \ge \delta _{ij} \frac{\hbar}{2}$.
As was for the $\tanh$ modification, the modified momentum uncertainty for the $\arctan$ modification has $\Delta P_j \le p_M$, which then gives the same minimum as for the $\tanh$ modification. Namely, $\Delta X^{min} _i = \frac{\hbar}{2 p_M}$.
The fact that this rough estimate for $\Delta X^{min} _i$ is the same for both $\tanh$ and $\arctan$ modification is not surprising. To this level of approximation, the right hand side of the commutator is the standard one, $i \hbar \delta _{ij}$. $\Delta X^{min} _i$ ultimately comes from the cut-off in the momentum uncertainty, which is the same in both cases, namely $\Delta P_j \le p_M$. 

To get a better estimate for $\Delta X^{min} _i$, we need a finer lower bound on $(1+x^2) \frac{\arctan (x)}{x}$. This is given by the function $(1 + x^2)^{1/2}$, {\it i.e.} $(1+x^2) \frac{\arctan (x)}{x} \ge (1+x^2)^{1/2}$. The fact that $(1 + x^2)^{1/2}$ provides a lower bound to $(1+x^2) \frac{\arctan (x)}{x}$ is not obvious (there are other possible functions that may provide a tighter lower bound), but can be verified by expanding both functions, or by graphing them. 

In any case, replacing $(1+x^2) \frac{\arctan (x)}{x}$ by $(1 + x^2)^{1/2}$ in \eqref{arctan-cond} and expanding to second order in $x = \frac{\pi |{\vec p}|}{2 p_M}$ gives 
\begin{equation}
    \label{arctan-comm-3}
     [{\hat X}_i , {\hat P}_j] \approx i \hbar     \left[ \delta_{ij}+ \frac{1}{2} \left( \frac{\pi |{\vec p}|}{2 p_M}\right)^2  \left( \delta_{ij} - \frac{p_j p_i}{|{\vec p}|^2} \right)  \right] ~.
\end{equation}
The commutator in \eqref{arctan-comm-3} yields the following uncertainty relationship
\begin{equation}
    \label{arctan-gup}
    \Delta X_i \Delta P_j \ge \frac{\hbar}{2} \left(\delta_{ij} + \frac{\delta _{ij} \pi ^2}{8 p_M ^2} \langle |{\vec p}|^2 - p_i ^2 \rangle  \right)  = \frac{\hbar}{2} \delta_{ij} \left( 1 + \frac{\pi ^2 \langle p_{\perp} ^2 \rangle}{8 p_M ^2}   \right) ~.
\end{equation}

In going from the \eqref{arctan-comm-3} to \eqref{arctan-gup} we have again used spherical symmetry to write the expectation of the last term in \eqref{arctan-comm-3} as $\frac{\pi ^2 \langle p_j p_i \rangle}{8 p_M ^2} = \delta _{ij} \frac{\pi ^2 \langle p_i ^2 \rangle}{8 p_M ^2}$, such that the expectation is zero unless $i=j$; and as before, we define $\langle p_{\perp} ^2 \rangle = \langle |{\vec p}|^2 - p_i ^2 \rangle$. From \eqref{arctan-gup} we now repeat all the steps going from \eqref{tanh-gup-2} to \eqref{dpj2} to obtain the following expressions for the $\arctan$ modification
\begin{equation}
    \label{arctan-up-5}
   \Delta X_i \ge   \frac{\hbar}{2} \left( \frac{1}{\Delta p_i} + \frac{\pi ^2 \Delta p_i}{8 p_M ^2}   \right)~,
\end{equation}
The minimum of \eqref{tanh-up-2} occurs at $\Delta p_i = p_M$ which then gives
\begin{equation}
    \label{arctan-xmin}
    \Delta X^{min} _i = \frac{\pi \hbar}{2\sqrt{2} p_M}. 
\end{equation} 

Equations \eqref{arctan-up-5} and \eqref{arctan-xmin} are the $\arctan$ modification analogs of \eqref{tanh-up-2} and \eqref{tanh-xmin}, only they differ by factors of order $1$. Notice, there is not that big a difference between the minima from \eqref{tanh-xmin} and \eqref{arctan-xmin}, and the rougher estimate obtained when the minimum in position uncertainty which came entirely from $\Delta P_i$ being bounded by $p_M$, where $\Delta X^{min}_i = \frac{\hbar}{2 p_M}$.

\subsection{Case with momentum in one direction}

We now look at the physical consequences on the modified uncertainty relationship when the system has some overall average momentum in one direction, {\it e.g.} ${\vec p} = (p_1, 0, 0)$. For this choice in momentum, the right hand side of \eqref{tanh-cond} and \eqref{arctan-cond} simply becomes $1$ when $i=j=1$, which gives the standard commutator $[X_1, P_1] = i \hbar$ for both cases. 
In the $1$ direction, the uncertainty relationship becomes $\Delta X_1 \Delta P_1 \ge \frac{\hbar}{2}$, which is just the standard uncertainty but constructed with modified operators.

For both $\tanh$ and $\arctan$ modifications, one has $\Delta P_1 \le p_M$; the same position uncertainty minimum in the $1$ direction of $\Delta X^{min}_1 \ge \frac{\hbar}{2 p_M}$. Note that this minimum in the $1$-direction is the same for both types of modified operators when the right hand side is bounded by the standard commutator.

We now turn to the uncertainty in the two orthogonal directions $2$ and $3$. For the modified operators $\hat{X}_j$ and $\hat{P}_j$ with $j=2$ or $3$ for the $\tanh$ modification, the commutator becomes
\begin{equation}
    \label{tanh-comm-2}
    [{\hat X}_{j} , {\hat P}_{j}] = i \hbar \left[ \frac{p_M}{|{\vec p}|} \sinh \left( \frac{|{\vec p}|}{p_M} \right) \cosh \left( \frac{|{\vec p}|}{p_M} \right)\right] = i\hbar\left[\cosh^2\left(\frac{|\vec{p}|}{p_M}\right) \frac{p_M}{|{\vec p}|} \tanh\left(\frac{|\vec{p}|}{p_M}\right) \right]~.
\end{equation}
The associated uncertainty is then 
\begin{equation}
    \label{tanh-up}
    \Delta X_j \Delta P_j \ge \frac{\hbar}{2} \left\langle \frac{p_M}{|{\vec p}|} \sinh \left( \frac{|{\vec p}|}{p_M} \right) \cosh \left( \frac{|{\vec p}|}{p_M} \right) \right\rangle \ge \frac{\hbar}{2} \left( 1 +  \frac{\langle |{\vec p}|^2 \rangle }{2 p_M^2}  \right) ~,
\end{equation}
where in the last step we expanded the hyperbolic functions to second order. For the $1$-direction modified momentum uncertainty, we take the rough estimate that the momentum uncertainties in the $2,3$-directions are bounded by $p_M$,  {\it i.e.} $\Delta P_j \le p_M$. Again, we assume $\langle p_j \rangle =0$ and $\langle P_j \rangle =0$. With this, the modified position uncertainty minimum is 
\begin{equation}
    \label{tanh-up-3}
    \Delta X^{min}_j = \frac{\hbar}{2 p_M} \left( 1 +  \frac{\langle |{\vec p}|^2 \rangle }{2 p_M^2}  \right) \ge \frac{\hbar}{2 p_M} \left( 1 +  \frac{ p_1 ^2}{2 p_M^2}  \right) ~.
\end{equation}
In the last expression in \eqref{tanh-up-3}, we used $\langle |{\vec p}|^2 \rangle \ge p_1 ^2$ since we assume that the momentum is all in the $1$-direction. This leads to the conclusion that the minimum uncertainty for the modified position in the $1$-direction is smaller than in the orthogonal, $2,3$-directions; such that
$\Delta X^{min}_{2,3} > \Delta X^{min}_1$.

We now repeat the above calculations for the $\arctan$ modification. In the $1$-direction one has as before $\Delta X^{min} _1 = \frac{\hbar}{2 p_M}$, by taking the rough bound $\Delta P_1 \le p_M$. Now for the two orthogonal directions, $j=2,3$ the commutator becomes
\begin{equation}
    \label{arctan-comm-2}
     [{\hat X}_j , {\hat P}_j] = i \hbar \left[ \left( 1+ \left( \frac{\pi |{\vec p}|}{2 p_M}\right)^2 \right) \frac{2 p_M}{\pi |{\vec p}|} \arctan \left( \frac{\pi |{\vec p}|}{2 p_M} \right) \right] ~.
\end{equation}
This commutator in \eqref{arctan-comm-2} now gives the modified uncertainty relationship 
for $j=2,3$
\begin{eqnarray}
    \label{arctan-up}
    \Delta X_j \Delta P_j &\ge& \frac{\hbar}{2}  \left\langle \left( 1+ \left( \frac{\pi |{\vec p}|}{2 p_M}\right)^2 \right)  \left(\frac{2p_M}{\pi |\vec{p}|} \right) \arctan \left( \frac{\pi |{\vec p}|}{2 p_M} \right)\right\rangle \nonumber \\
    &\ge& 
    \frac{\hbar}{2}  \left\langle \sqrt{\left( 1+ \left( \frac{\pi |{\vec p}|}{2 p_M}\right)^2 \right)}  \right\rangle \ge \frac{\hbar}{2} \left( 1+ \frac{\pi^2 \langle |{\vec p}|^2 \rangle}{8 p_M} \right)~.
\end{eqnarray}

In going from the first line to the second line in \eqref{arctan-up}, we have employed the previously used non-obvious lower bound $(1+x^2) \frac{\arctan(x)}{x} \ge \sqrt {1 +x^2}$, and taken a Taylor expansion $\sqrt{1+x^2} \approx 1 + \frac{1}{2} x^2$. If we again take the rough bound of $\Delta P_j \le p_M$ for the modified momentum in the two orthogonal directions we obtain a minimum in the modified position uncertainty of 
\begin{equation}
    \label{arctan-up-2}
    \Delta X ^{min} _j  = \frac{\hbar }{2p_M}  \left( 1+ \frac{\pi^2 \langle |{\vec p}|^2 \rangle}{8 p_M} \right) \ge 
    \frac{\hbar }{2p_M}  \left( 1+ \frac{\pi^2 p_1 ^2}{8 p_M} \right)~.
\end{equation}
Again, we have used $\langle |{\vec p}|^2 \rangle \ge p_1 ^2$ in arriving at the last expression in \eqref{arctan-up-2}. 

As for the $\tanh$ modification, the term in parentheses in \eqref{arctan-up-2} is greater than $1$. So, as before, the uncertainty in the orthogonal $2,3$-directions is larger than in the $1$-direction - the direction of the momentum of the system. This can be compared to when the electric field of a boosted, charged particle is stronger in the directions orthogonal to the momentum of the particle.

\section{Summary and conclusions}

In this paper we formulated 3D GUP models based on earlier 1D GUP models studied in \cite{aiken-2019,BLS,BJLS}. Our aim in 1D was to find a GUP model which gave a minimal length which only modified the position and momentum operators, while leaving the commutator of these modified operators unmodified. 
In section II, we summarized the 1D GUP models. In section III, we extended these 1D models to 3D GUP models so that they were rotationally invariant (coordinate-independent) by requiring the modifications to be functions of the total momentum $|\vec{p}|$.  
Consequently, the 3D commutator must be modified in at least one direction. 
This can be seen explicitly in \eqref{tanh-comm} and \eqref{arctan-comm} via the second order corrections to the the position-momentum commutators which had a form $\propto \delta_{ij} - \frac{p_j p_i}{|{\vec p}|^2}$. In 1D GUP models, the term $\delta_{ij} - \frac{p_j p_i}{|{\vec p}|^2}$ is always zero. 
We then investigated the uncertainty relationships produced by the modified 3D operators and modified commutators. 
In general, we found that the minimum uncertainty in position was approximately $\Delta X_i ^{min} \sim \frac{\hbar}{p_M}$  - essentially the same result as in 1D. 

We also considered the minimum distance in two cases.
In the first case, we used spherically-symmetric wave functions to analyze how small of a volume the wave function could be compressed; we found that it was the minimal distance cubed, up to a factor of order one.  
In the second case, we considered a system with a large, non-zero momentum in some particular direction. Here we found that the minimum distance in the direction of the average momentum was different from the minimum distance in the orthogonal directions. For either the $\tanh$ and $\arctan$ GUPs, the positional uncertainty in the orthogonal directions was larger than in the direction of the non-zero momentum. A similar result occurs in the GUP model in reference \cite{KMM} {\it i.e} $\Delta X_{2,3} ^{min} > \Delta X_1 ^{min}$.

The difference between the minimum position uncertainties in the second case above hints that there may be a violation of Lorentz symmetry. 
Moreover, the existence of minimal distances would naively imply violation of Lorentz symmetry. 
However in \cite{BJLS}, a GUP model was formulated which gave a minimal length and preserved the special relativistic energy and momentum relationship, $E^2 - |{\vec p}|^2 = m^2$, and thus preserved some aspects of the Lorentz symmetry.  
A complementary issue is the connection of our work to a breaking of the isotropy of space, as observed for other GUP models \cite{husin}. In future work, we plan to study the issue of the fate of the Lorentz symmetry of these modified position and momentum operators. 

The standard Lorentz algebra is 
generated by the angular momentum operators and boost operators in terms of the time-position operators and energy-momentum operators.  
One major question to address is how the modified position and momentum operators affect the observables associated with modified angular momentum and boost operators. 
To obtain the modified boost operators we will need to also define modified time and energy operators that fit in with the $\tanh$ or $\arctan$ operators of equations \eqref{xp3dc} and \eqref{xp3dd}. 

Another related issue is the modification of commutation relations of position and momentum operators among themselves. 
The standard result is $[{\hat x}_i, {\hat x}_j] =  0$ and $[{\hat p}_i, {\hat p}_j]=  0$. 
While this is still true for the modified momentum operators \eqref{xp3dc} and \eqref{xp3dd}, it is no longer true for the modified position operators {\it i.e.}, $[{\hat P}_i, {\hat P}_j]=  0$ but $[{\hat X}_i, {\hat X}_j] \ne  0$. 
This non-commutativity of the modified position operators provides a potential link between GUP models and non-commutative geometry models \cite{piero,euro} which is another approach to minimal lengths.

Also, the modified position operators from \eqref{xp3dc} and \eqref{xp3dd} satisfy $[{\hat X}_i, {\hat X}_j] \propto \epsilon_{ijk}  {\hat l}_k$, so that the commutators are proportional to the angular momentum operators. 
For the related uncertainty relationships, this implies that $\Delta X_i \Delta X_j \propto  \epsilon_{ijk}  \langle  {\hat l}_k\rangle $; the uncertainty in area is related to the expectation of the angular momentum operator. 
This provides a potential connection between the 3D GUP models in this work and spin  foam models \cite{baez,alejandro} which have a similar connection between area and angular momentum. 

We leave both of these questions -- the modification of the Lorentz algebra and the connection of 3D GUP models to non-commutative geometry and spin foam models -- for a subsequent, companion paper. \\

{\bf Acknowledgment:} DS is a 2023-2024 KITP Fellow at the Kavli Institute for Theoretical Physics and his work was partially supported in part by the National Science Foundation under Grant No. NSF PHY-1748958. The work of PN is  partially been supported by GNFM, Italy’s National Group for Mathematical Physics. The work of MB and DS were supported through a Fresno State 2023-2024 RSCA grant.

\end{document}